\documentclass[12pt]{article}
\usepackage{amsmath, amssymb}

\newtheorem{lemma}{Lemma}[section]

\newtheorem{theorem}{Theorem}[section]

\newtheorem{proposition}{Proposition}[section]

\def \ddt {\frac{d}{dt}}

\def \DD  {{\cal D}}

\def \FF  {{\cal F}}

\def \LL  {{\cal L}}

\def \NN  {{\cal N}}

\def \OO  {{\cal O}}

\def \TT  {{\cal T}}

\def \UU  {{\cal U}}

\def \WW  {{\cal W}}

\def \CCC {\mathbb{C}}

\def \EEE {\mathbb{E}}

\def \FFF {\mathbb{F}}

\def \HHH {\mathbb{H}}

\def \KKK {\mathbb{K}}

\def \NNN {\mathbb{N}}

\def \PPP {\mathbb{P}}

\def \RRR {\mathbb{R}}

\def \TR {\hbox{Tr}}

\def \dG  {{\mathrm d}\Gamma}

\def \doverbar  {{\mathrm d}\overline{\Gamma}}

\title{Accuracy of the time-dependent Hartree-Fock approximation}

\author{
Claude BARDOS%
\footnote{ University of Paris 7  and LAN (Univ. Paris 6), France
(bardos@math.jussieu.fr). },
Fran\c cois GOLSE%
\footnote{ ENS-Ulm and LAN (Univ. Paris 6), France
(Francois.Golse@ens.fr). },
Alex D.\  GOTTLIEB %
\footnote{ Wolfgang Pauli Inst. c/o Inst.\ f.\ Mathematik, Univ.
Wien, Strudlhofg.\ 4, A--1090 Wien, Austria
(alex@alexgottlieb.com).
} \\
and
Norbert J.\ MAUSER%
\footnote{ Wolfgang Pauli Inst. c/o  Inst.\ f.\ Mathematik, Univ.
Wien, Strudlhofg.\ 4, A--1090 Wien, Austria
(mauser@courant.nyu.edu). } }

\date{ }

\begin{document}

\maketitle

\begin{abstract}
This article examines the time-dependent Hartree-Fock (TDHF)
approximation of single-particle dynamics in systems of
interacting fermions.   We find the TDHF approximation to be
accurate when there are sufficiently many particles and the
initial many-particle state is a Slater determinant, or any Gibbs
equilibrium state for noninteracting fermions.  Assuming a bounded
two-particle interaction, we obtain a bound on the error of the
TDHF approximation, valid for short times.  We further show that
the error of the the TDHF approximation vanishes at all times in
the mean field limit.
\end{abstract}

\section{Introduction}

  Dirac \cite{Dirac} invented the time-dependent Hartree-Fock
equation in 1930.  The time-dependent Hartree-Fock (TDHF) equation
is a nonlinear Schr\"odinger equation designed to approximate the
evolution of the single-electron state in an $n$-electron system.
Dirac noted that the TDHF equation, originally written as a system
of $n$ coupled Schr\"odinger equations for occupied orbitals, may
be written as a Liouville-von Neumann equation for the
single-particle reduced density operator.  We study the TDHF
equation in this form, availing ourselves of trace norm
techniques inspired by \cite{Spohn} to estimate the error of the
TDHF approximation.

We will consider Hamiltonian dynamics of fermions interacting
through a two body potential $V$. The energy operator for a
solitary particle will be denoted by $L$, the energy of
interaction of a single pair of particles will be denoted by $V$,
and the total energy operator for a system of particles will be
the sum of all single-particle energies and all pair energies. In
this article we only consider bounded interaction potentials $V$;
the case where $V$ represents Coulomb repulsion between electrons
will be treated in a forthcoming paper \cite{BGGM2}. Although the
number of particles does not change under the dynamics just
described, we prefer to formulate the dynamics on a fermion Fock
space, so that we may consider initial states of indeterminate
particle number.  We are particularly interested in initial states
which are Gibbs equilibrium states (grand-canonical ensembles),
for we are going to show that such initial states enhance the
accuracy of the TDHF approximation.

 Let $L$ be a
self-adjoint operator on $\HHH$, and let $V$ be a bounded
Hermitian operator on $\HHH \otimes \HHH$ that commutes with the
transposition operator $U$ defined by $U(x\otimes y)=y\otimes x$.
We are going to discuss the dynamics on the fermion Fock space
$\FFF_{\HHH}$ whose Hamiltonian $H$ may be written in second
quantized form as
\begin{equation}
\label{Ham}
       H \ = \ \sum_{i,j} \langle j| L| i \rangle a^{\dagger}_j a_i
       \ + \ \sum_{i,j,k,l} \langle k l | V | i j \rangle
       a^{\dagger}_k
       a^{\dagger}_l a_j a_i
\end{equation}
   We will analyze the solutions of the Liouville-von
Neumann equation
\begin{eqnarray}
         i \hbar \frac{d}{dt} D(t) & = & [H,D(t)]   \nonumber \\
               D(0) & = & D_0
         \label{vN},
\end{eqnarray}
which is the evolution equation for the density operator in the
 Schr\"odinger picture of quantum dynamics.

 We will see that (\ref{vN}) leads to the following equation for
the single-particle density operator $\NN_1(t)$:
\begin{eqnarray}
         i \hbar \frac{d}{dt} \NN_1(t) & = & [L,\NN_1(t)] \ + \ [V, \NN_2(t) ]_{:1}
     \nonumber  \\
                              \NN_1(0) & = & \NN_1(D_0).
     \label{vN1}
\end{eqnarray}
(The subscript $_{:1}$ denotes a partial trace; see definition
(\ref{partialTraceFormula}) below.) This equation for $\NN_1(t)$
is not ``closed," for its right hand side involves the
two-particle density operator $\NN_2(t)$.
 The TDHF approximation to $\NN_1(t)$ is the solution of the initial value problem
\begin{eqnarray}
  i\hbar \ddt F(t) & = & [L,F(t)] \ + \ [V,\ F(t)^{\otimes 2}2A_2]_{:1} \nonumber  \\
  F(0) & = & \NN_1(D(0))
  \label{TDHF}
\end{eqnarray}
where $A_2$ is the orthogonal projector of $\HHH\otimes \HHH$ onto
the subspace of antisymmetric vectors.  The existence and
uniqueness of solutions of (\ref{TDHF}) were established in
\cite{Bove1} for the case where $V$ is a bounded operator, and in
\cite{Chadam, Bove2} for the case where $V$ is a Coulombic
interaction.

The TDHF equation (\ref{TDHF}) is obtained by closing the
single-particle equation (\ref{vN1}) with the Ansatz
\begin{equation}
\label{Ansatz}
        \NN_2 = (\NN_1\otimes \NN_1)2A_2
\end{equation}
at all times.  The relation (\ref{Ansatz}) holds for pure states
corresponding to Slater determinants, and also for Gibbs
densities.  However, even supposing that $\NN_2(0)$ satisfies
(\ref{Ansatz}), the interaction $V$ is likely to introduce
``correlations" in $\NN_2(t)$, that is, departures from
(\ref{Ansatz}), and ignoring those correlations in the TDHF
equation requires some justification.

We are going to prove that the absence of correlations is
self-perpetuating in the mean field limit: if $\NN_2(0)$ satisfies
(\ref{Ansatz}) then $\NN_2(t)$ asymptotically satisfies
(\ref{Ansatz}) as the number of particles $N$ tends to infinity
and the interaction strength is scaled as $1/N$.  This is the
content of Theorem~\ref{mean-field-theorem}.
To prove this theorem we first bound the error of the TDHF
approximation in terms of the average particle number and interaction strength.
 This bound is presented in Theorem~\ref{short-times} without
reference to the mean field scaling, and it applies to any
system of fermions whose Hamiltonian has the form (\ref{Ham}).  Unfortunately,
the bound is valid only at short times, i.e., up to a time inversely
proportional to $\|V\|$ and the average particle number.

Let us advert to some shortcomings of our results. Firstly, we
consider only bounded two-particle potentials $V$. Fortunately,
the challenges presented by the Coulomb potential can be overcome
\cite{BGGM2, BEGMY} and the mean field limit of
Theorem~\ref{mean-field-theorem} also holds for certain electronic
systems.  Secondly, the explicit error bound of
Theorem~\ref{short-times} is valid only at short times, which are
too short to be of interest in molecular-electronic problems, even
if the Coulomb interaction between electrons is truncated at the
Bohr radius. However, the model (\ref{Ham}) does not only apply to
molecular-electronic problems, and we hope that
Theorem~\ref{short-times} may find other applications, perhaps to
certain models of interactions between nucleons \cite{BDGGM}.

We first published our derivation of the TDHF equation in the mean
field limit in \cite{BGGM1}.  There we assumed that the initial
states are Slater determinants. The main theme of this article is
that the initial states need not be Slater determinants; the TDHF
approximation should work equally well (or badly) for all initial
many-fermion states of Gibbs type.   Also, in this article we are
not only interested in the mean field limit, and we offer the
error bound of Theorem~\ref{short-times} for the unscaled problem.

 The next section gives the background on fermion Fock space,
trace class operators, reduced density operators, and Gibbs
equilibrium states. Section~\ref{carefully} carefully defines the
many-particle dynamics generated by (\ref{Ham}) and derives the
single-particle equation (\ref{vN1}) for $\NN_1(t)$.
Section~\ref{hartreehierarchy} introduces the TDHF approximation
of $\NN_1(t)$ and its higher-order analogs. Section~\ref{accuracy}
presents Theorems~\ref{short-times} and \ref{mean-field-theorem}.
Finally, the appendix contains the proofs of several Propositions.




\section{Definitions and notation}
\label{notation}

 Consider a quantum particle whose Hilbert space is $\HHH$, i.e.,
 a particle which, in isolation, would constitute
a system whose (pure) quantum states are represented by the
rank-one orthogonal projectors on some Hilbert space $\HHH$. The
set of quantum states available to a system of $n$ particles of
this kind is determined by their ``statistics," i.e., whether the
particles are fermions, bosons, or distinguishable.  If the
particles are fermions, the pure states of a system of $n$ of them
are represented by the rank-one projectors onto vectors in the
{\it antisymmetric subspace} $\HHH^{(n)}$ of the tensor power
space $\HHH^{\otimes n}$.

 The Hilbert space $\HHH^{\otimes n}$ is the
closed span of the simple tensors
\[
   x_1 \otimes x_2 \otimes \cdots \otimes x_n \qquad \qquad
   x_1,x_2,\ldots,x_n \in \HHH
\]
with the inner product
\[
     \langle y_1 \otimes y_2 \otimes \cdots \otimes y_n, \ x_1 \otimes x_2 \otimes \cdots \otimes
     x_n\rangle \ = \ \langle y_1, x_1 \rangle \langle y_2, x_2
     \rangle \cdots \langle y_n, x_n \rangle.
\]
Let $\Pi_n$ denote the group of permutations of
$\{1,2,\ldots,n\}$.  For each $\pi \in \Pi_n$, a unitary operator
$U_{\pi}$ on $\HHH^{\otimes n}$ may be defined by extending
\[
        U_{\pi}(x_1 \otimes x_2 \otimes \cdots \otimes x_n) \ = \
        x_{\pi^{-1}(1)} \otimes x_{\pi^{-1}(2)} \otimes \cdots \otimes x_{\pi^{-1}(n)}
\]
to all of $\HHH^{\otimes n}$.  A vector $\psi \in \HHH^{\otimes
n}$ is {\it antisymmetric} if $U_{\pi}(\psi) =
\hbox{sgn}(\pi)\psi$ for all $\pi \in \Pi_n$.  The antisymmetric
vectors in $\HHH^{\otimes n}$ form a closed subspace which will be
denoted $\HHH^{(n)}$. The orthogonal projector with range
$\HHH^{(n)}$ is
\begin{equation}
\label{An}
    A_n \ = \ \frac{1}{n!}\sum_{\pi \in \Pi_n} \hbox{sgn}(\pi)
    U_{\pi}.
\end{equation}
If $x_1,x_2,\ldots,x_n$ is an orthonormal system in a Hilbert
space $\HHH$, then the vector
\begin{equation}
\label{SlaterD}
       \sqrt{n!} \ A_n(x_1 \otimes x_2 \otimes \cdots \otimes
     x_n),
\end{equation}
is a unit vector in $\HHH^{(n)}$.  A vector of the form
(\ref{SlaterD}) is called a {\it Slater determinant}.  If
$\{e_{\alpha}\}$ is a complete orthonormal system in $\HHH$, then
a basis of $\HHH^{(n)}$ may be chosen from the set of all Slater
determinants of the form (\ref{SlaterD}) where
$\{x_1,\ldots,x_n\}$ a subset of the single particle basis of
cardinality $n$.

If the number of particles in the system is not fixed, the
appropriate Hilbert space is the direct sum of the $n$-particle
spaces $\HHH^{(n)}$. This is the (fermion) Fock space
\begin{equation}
\label{FockSpace}
     \FFF_{\HHH} \ = \   \HHH^{(0)} \oplus \HHH^{(1)} \oplus \HHH^{(2)} \oplus \HHH^{(3)} \oplus \cdots \qquad .
\end{equation}
The possibility of a zero-particle state is accommodated by
$\HHH^{(0)} \approx \CCC $, a one-dimensional space spanned by the
{\it vacuum vector} $\Omega$.  The {\it number operator} $N$ on
$\FFF_{\HHH}$ is the self-adjoint operator whose restriction to
$\HHH^{(n)}$ equals multiplication by $n$.  Annihilation and
creation operators $a_x$ and $a^{\dagger}_x$ are explicitly
represented on $\FFF_{\HHH}$ as discussed in \cite{Bratteli}. The
statistical state of a system of fermions with single-particle
space $\HHH$ determines a bounded positive continuous functional
$\omega$ on the bounded operators on $\FFF_{\HHH}$ with
$\omega(I)=1$.  We are only going to consider {\it normal states},
i.e., those $\omega$ that satisfy
\[
      \omega(A) \ = \ \TR(D_{\omega}A)
\]
for some nonnegative trace class operator $D_{\omega}$ of unit
trace, which may be called the {\it (statistical) density
operator}.

Density operators on any Hilbert space $\KKK$ are trace class
operators in particular.  Let $\TT(\KKK)$ denote the real Banach
space of Hermitian trace class operators $T$ with the norm $
\|T\|_1 \ = \ \TR(|T|)$. We often use the inequality that, for
bounded operators $B$ on $\KKK$,
\begin{equation}
\label{basicInequality}
        |  \TR( T B ) | \     \le \     \|T \|_1\|B\|.
\end{equation}
A linear functional defined on the space of compact Hermitian
operators by
\begin{equation}
\label{duality}
      \Phi_T(K) \ = \   \TR( T K )
\end{equation}
is continuous by (\ref{basicInequality}), and indeed the space
$\TT(\KKK)$ is isomorphic to the Banach dual of the space of
compact Hermitian operators on $\KKK$, via the isomorphism  $T
\longmapsto \Phi_T$.


\subsection{Reduced density operators}

We now restrict our attention to density operators $D$ on
$\FFF_{\HHH}$ that commute with the number operator and such that
$N^mD$ is trace class for all $m \in \NNN$. For such $D$ we will
define the $ m^{th}$ {\it order reduced number density operators}
$\NN_m(D)$ and explain their physical significance.

We begin by defining the reduction of an $n$-particle statistical
density operator, i.e., a positive operator of unit trace $D$ on
$\HHH^{(n)}$, to an $m$-particle density operator denoted
$D_{:m}$.
 If $A$ is an operator on $\HHH^{(n)}$, let $\overline{A}$ denote
the extension of $A$ to $\HHH^{\otimes n}$ defined by first
projecting onto $\HHH^{(n)}$ and then applying $A$, i.e.,
\[
     \overline{A} \ = \ A
     \oplus
     {\bf 0}_{ \HHH^{\otimes n} \ominus \HHH^{(n)} }.
\]
 For each $0\le m \le n$ there exists a positive contraction
from $\TT(\HHH^{(n)})$ onto $\TT(\HHH^{(m)})$ known as the {\it
partial trace}. This partial trace map, which we denote by $T
\mapsto T_{:m}$, is defined implicitly through the duality
(\ref{duality}) by the requirement that
\[
     \TR( (\overline{K} \otimes I \otimes \cdots \otimes I) \overline{T}) \ = \ \TR(K
     T_{:m})
\]
for all compact Hermitian operators $K$ on $\HHH^{(m)}$.  It
follows that, for any orthonormal basis $\OO$ of $\HHH^{\otimes
n-m}$,
\begin{equation}
\label{partialTraceFormula}
      \langle \xi, T_{:m} \psi \rangle
      \ = \ \sum_{\phi \in \OO}
      \langle (\xi \oplus 0) \otimes \phi , \ \overline{T}((\psi\oplus 0) \otimes \phi) \rangle \ .
\end{equation}
In case $D$ is a density operator on $\HHH^{(n)}$, and if $m \le
n$, the operator $D_{:m}$ is known as the {\it m-particle reduced
density operator} \cite{TerHaar}.  It is used to determine the
expected values of $m$-particle observables when the $n$-particle
system is in the statistical state $D$. The $m$-particle reduced
density operators obtained from $P_{\psi}$ will be denoted
$\left(P_{\psi}\right)_{:m}$.  If $\psi$ is an $n$-particle Slater
determinant, then the reduced density operators
$\left(P_{\psi}\right)_{:m}$ satisfy
\begin{eqnarray}
  \left(P_{\psi}\right)_{:1} & = &  \frac{1}{n}\sum_{j=1}^n P_{x_j}
  \label{SlaterOne}  \\
   \left(P_{\psi}\right)_{:m} & = &  \frac{n^m}{\binom{n}{m}}
  \left(P_{\psi}\right)_{:1}^{\otimes m} A_m .
  \label{SlaterFormula}
\end{eqnarray}

Now we can define the reduced number densities $\NN_m$, which serve to
describe the $m$-particle correlations in a system of many particles.
  Let $D$
be a density operator that commutes with $N$.  Then
\begin{equation}
\label{directsum}
    D \ = \ \bigoplus_{n=0}^{\infty} D_n
\end{equation}
where each $D_n$ is a nonnegative trace class operator on
$\HHH^{(n)}$.  Assume that
\begin{equation}
\label{m-moment}
    \sum_{n=0}^{\infty} n^m \TR(D_n) \ < \ \infty.
\end{equation}
For such $D$, define the $m^{th}$ order reduced number density
\begin{equation}
\label{sumofmarginals}
     \NN_m(D) \ = \  \sum_{n=m}^{\infty}\frac{n!}{(n-m)!} D_{n:m}.
\end{equation}
The operators $\NN_m(D)$ are called
{\it reduced density operators } in Section 6.3.3 of \cite{Bratteli},
but we prefer to call them reduced {\it number} density
operators because the trace of $\NN_1$ is the average particle number.
They are trace class by (\ref{m-moment}).
An $n$-particle density operator $D_n$ extends
to a density operator
\begin{equation}
\label{extend}
  \widehat{D_n} \ = \ {\bf 0} \oplus \cdots \oplus {\bf 0} \otimes D_n \oplus {\bf 0}
  \oplus \cdots
\end{equation}
on all of Fock space.  It is clear that $\NN_m(\widehat{D_n})$
equals $
     \NN_m(\widehat{D_n}) = n(n-1)\cdots(n-m+1) D_{n:m}
$ if $m \le n$ but it equals the zero operator if $m>n$.

Formula (\ref{significanceOfN1}) below
may clarify the sense in which $\NN_1$ determines expected values of
single-particle observables. Following \cite{Bratteli}, we define
the second quantization $\dG(H)$ of a self-adjoint operator $H$ on
$\HHH$ to be the closure of the essentially self-adjoint operator
$\oplus \dG_n(H)$ on $\FFF_{\HHH}$, where $\dG_n(H)$ denotes the
restriction to $\HHH^{(n)}$ of the operator
\begin{equation}
     \doverbar_n(H) \ = \  \sum_{j=1}^n
      \stackrel{j-1  \ \  times}{I \otimes \cdots \otimes I}
    \otimes H \otimes
    \stackrel{n-j  \ \  times}{I \otimes \cdots \otimes I}
    \label{doverbar}
\end{equation}
on $\HHH^{\otimes n}$.  The operator $\dG(H)$ is called the {\it
second quantization of} $H$; we might also call it a {\it
single-particle observable}.    Suppose that $\omega$ is a normal
state on $\FFF_{\HHH}$ whose density $D_{\omega}$ satisfies
(\ref{directsum}) and (\ref{m-moment}).   From
(\ref{sumofmarginals}) it then follows that
\begin{equation}
\label{significanceOfN1}
      \omega(\dG(B)) \ = \  \TR(\NN_1(D_{\omega}) B)
\end{equation}
for any bounded Hermitian operator $B$ on $\HHH$.  For example,
the number operator $N$ on $\FFF_{\HHH}$ is the second
quantization of the identity operator, i.e., $N=\dG(I)$.  From
(\ref{significanceOfN1}) we see that the trace of
$\NN_1(D_{\omega})$ is the average number of particles in the
system when it is in the state $\omega$.  Another example: if $x$
is a unit vector in $\HHH$, the operator $a^{\dagger}_x a_x$ on
$\FFF_{\HHH}$ is the second quantization $\dG(P_x)$ of the
rank-one projector $P_x$, and (\ref{significanceOfN1}) tells us
that $\omega(a^{\dagger}_x a_x ) =  \TR(\NN_1(D_{\omega}) P_x)$.

Theorems~\ref{short-times} and \ref{mean-field-theorem} rely on
the following fact, which is proved in the appendix:
\begin{proposition}
\label{N1} If $D$ is a density operator on $\FFF_{\HHH}$ that
commutes with $N$ and such that $\TR(ND)<\infty$, then the
operator norm of $\NN_1(D)$ is not greater than $1$.
\end{proposition}


\subsection{Gibbs equilibrium states}
\label{gibbs}

A Gibbs equilibrium state is that of a system  of noninteracting
fermions in thermal and chemical equilibrium with its environment,
but only in case $e^{-\beta H}$ is trace class, where $H$ is the
single-particle Hamiltonian and $1/\beta$ is Boltzmann's constant
times the temperature.   In this case the density operator is
proportional to $\exp(-\beta \  \dG(H-\mu I))$, where $\mu \in
\RRR$ is the chemical potential.   This density operator is
diagonalizable with respect to an occupation number basis of Fock
space, and occupation numbers are stochastically independent. If
the single-particle Hamiltonian has discrete eigenvalues
$\epsilon_1 \le \epsilon_2 \le \cdots $, the probability of
occupation of the $j^{th}$ level equals
$1/(1+e^{\beta(\epsilon_j-\mu)})$.

When $D$ is a Gibbs equilibrium density then
$
      \NN_m(D) = \NN_1(D)^{\otimes m}m! A_m.
$
The same is true when $D$ is a pure state corresponding to a
Slater determinant.  Indeed, this relation characterizes the
``gauge-invariant quasifree states of the CAR algebra" that have a
trace class single-particle operator, and both Slater densities
and Gibbs equilibrium densities are of this type.

We describe these states in probabilistic language.
 Let $\{\phi_j\}_{j \in J}$ be a basis of $\HHH$. One basis of
$\FFF_{\HHH}$ consists of the vacuum vector $\Omega$ and all
Slater determinants made of vectors from $\{\phi_j\}$. This basis
is indexed by $\FF$, the set of finite subsets of $J$ including
the empty set. A nonempty subset ${\bf s}=\{ j_1,\ldots, j_n\}$ of
$J$ corresponds to a Slater determinant $\Psi(\bf{s})$ formed from
the vectors $\phi_{j_1},\ldots,\phi_{j_n}$.  (There are in fact
two such Slater determinants, opposite in sign, and we choose one
of them.) The empty subset of $J$ corresponds to the vacuum vector
$\Omega$, i.e., $P(\oslash)= \Omega$. Let $\PPP$ be a probability
measure on $\FF$ with the $\sigma$-field of all its subsets.  That
is, let $\PPP:\FF \longrightarrow [0,1]$ be equal to $0$ except on
a countable subset of $\FF$, on which subset $\sum \PPP({\bf s}) =
1$.
 Define the
random variables $N_j$ on $\FF$ by
\[
      N_j({\bf s}) \ = \ \Big\{
      \begin{array}{rcr}
           1 \qquad & \hbox{if} \ & j \in {\bf s} \\
           0 \qquad & \hbox{if} \ & j \notin {\bf s} \\
      \end{array}
\]
and define $N=\sum N_j$.  Suppose that the $N_j$ are independent
with respect to $\PPP$, and define
\begin{equation}
\label{Prob}
            \PPP(N_j = 1) \ = \ p(j).
\end{equation}
Then \begin{equation} \label{useMe}
 \sum_{{\bf s} \ni j_1,\ldots,j_n }
       \PPP({\bf s}) \ = \
       \PPP((N_{j_1 }=1) \cap \cdots \cap (N_{j_n}=1) )
       \ = \ \prod_{i=1}^n p (j_i)
\end{equation}
when $j_1,\ldots, j_n$ are distinct. Note that
$
             \EEE(N) \ = \ \sum\limits_{j \in J} p(j) \ < \ \infty
$
since $\PPP(N < \infty)=1$.

 Given a probability measure $\PPP$ on $\FF$, we define a density
operator on $\FFF_{\HHH}$ by
\begin{equation}
\label{G}
     G[\PPP ] \ = \
     \sum_{{\bf s} \in \FF }
      \PPP ({\bf s}) P_{\bf s}\ ,
\end{equation}
where $P_{{\bf s}}$ denotes the orthogonal projector onto the span
of $\Psi({\bf s})$.
\begin{proposition}
\label{Prop1}
 Let $\PPP$ be as in (\ref{Prob}) and (\ref{useMe}).
  Let $G$ denote the density operator
$G[\PPP]$ of (\ref{G}). Then for all $n \in \NNN$
\begin{equation}
\label{exactClosure}
      \NN_n(G) \ = \  \NN_1(G)^{\otimes n}n!A_n.
\end{equation}
\end{proposition}
The preceding proposition is proved in the appendix.
Proposition~\ref{Prop1} and the following proposition (also proved
in the appendix) together imply Proposition~\ref{Prop2} below:
\begin{proposition}
\label{Prop1.5} If $T$ is a Hermitian trace class operator then
\begin{equation}
\label{proveMe}
       \big\| T^{\otimes n}n!A_n  \big\|_1 \ \le     \big\| T
       \big\|_1^n.
\end{equation}
\end{proposition}

\begin{proposition}
\label{Prop2}
  Let $G$ be as in Proposition~\ref{Prop1}. Then
$
           \big\|\NN_n(G)\big\|_1 \ \le \  \big\| \NN_1(G)\big\|_1^n
$.
\end{proposition}




\section{Definition of the dynamics on Fock space}
\label{carefully}

We now define the dynamics (\ref{vN}) more carefully, and derive
the reduced dynamics (\ref{vN1}) from (\ref{vN}).

Let $H$ be a self-adjoint operator on $\HHH$, and let $V$ be a
bounded Hermitian operator on $\HHH\otimes \HHH$ that commutes
with the transposition operator $U(x\otimes y)=y\otimes x$. For $1
\le j \le n$, let $L_j$ denote the operator
\[
      \stackrel{j-1  \ \  times}{I \otimes \cdots \otimes I}
    \otimes L \otimes
    \stackrel{n-j  \ \  times}{I \otimes \cdots \otimes I}
\]
on $\HHH^{\otimes n}$ (the value of $n\ge j$ is not explicit in
the notation $L_j$ but it will always be clear from context). For
$1 \le i < j \le n$, let $U_{(ij)}$ denote the permutation
operator on $\HHH^{\otimes n}$ that transposes the $i^{th}$ and
$j^{th}$ factors of any simple tensor $x_1 \otimes \cdots \otimes
x_n$, and let
\[
    V_{ij}\ = \  U_{(1i)}U_{(2j)} \left( V \otimes I^{\otimes n-2}
    \right) U_{(2j)}U_{(1i)}
\]
(again, the domain $\HHH^{\otimes n}$ of $V_{ij}$ will always be
clear from context).
 For each $n$, define the operators
\begin{eqnarray*}
   L^{(n)} & = & \sum\limits_{j=1}^n L_j  \\
   H^{(n)} & = & L^{(n)} \ + \ \sum\limits_{1\le i < j \le n} V_{ij}
\end{eqnarray*}
on $\HHH^{(n)}$ (these operators are defined on all of
$\HHH^{\otimes n}$ but we are only considering their restrictions
to the invariant subspace $\HHH^{(n)}$). The Hamiltonian operator
$H$, which we had formally represented above by (\ref{Ham}), is
the direct sum $H = \bigoplus H^{(n)}$ defined on the domain
\[
    \DD(H) \ = \ \Big\{
    x = \oplus x_n \in \FFF_{\HHH}: \ \sum_n \big\| H^{(n)}x_n \big\|^2 \ < \
    \infty
    \Big\}.
\]
This operator is closed and self-adjoint (see Section 6.3.1 of
\cite{Bratteli}), and $-\frac{i}{\hbar}H$ is the generator of the
strongly continuous group
\[
    W_t \ = \  \bigoplus_{n=1}^{\infty} W^{(n)}_t
\]
of unitary operators on $\FFF_{\HHH}$, where $W^{(n)}_t =
\exp\big(-\frac{it}{\hbar}H^{(n)}\big)$.

It is convenient to have some notation for the free part of the
dynamics, and we will subsequently use
\[
   U^{(n)}_t \ = \  \exp\big( -\frac{it}{\hbar}L^{(n)}\big)
   \qquad\quad \hbox{ and } \qquad\quad
     U_t  \ = \    \bigoplus_{n=1}^{\infty} U^{(n)}_t.
\]

 The Liouville-von Neumann dynamics corresponding to (\ref{vN}) are given by the
group
\begin{equation}
     \WW_t(D) \ = \    W_t D W_{-t}
\label{dynamics}
\end{equation}
of isometries of $\TT(\FFF_{\HHH})$. (See Proposition 3.4 of
\cite{Bove1} for a proof that groups of isometries defined in this
way are strongly continuous.)  Define the subspaces $\TT_n \subset
\TT(\HHH)$ consisting of all $n$-particle trace class operators:
\[
    \TT_n \ = \  \left\{ \widehat{T} \hbox{\ as in (\ref{extend})} \ : \  T_n \in \TT(\HHH^{(n)})\right\}.
\]
These subspaces are invariant under $\WW_t$, and the restriction
of $\WW_t$ to $\TT_n$ is
\[
     \WW^{(n)}_t(T) \ = \    W^{(n)}_t T W^{(n)}_{-t}.
\]
The generator of this group of isometries is
\begin{equation}
\label{generator}
   \LL_n
   \ + \  \sum_{1 \le i< j \le n} \left[V_{ij},\ \cdot \ \right],
\end{equation}
where $\LL^{(n)}$ is the the generator of the group
\begin{equation}
\label{noninteracting}
     \UU^{(n)}_t(T) \ = \    U^{(n)}_t T U^{(n)}_{-t},
\end{equation}
which may be written formally as $-\frac{i}{\hbar}\sum\limits^n
\left[L_j,\ \cdot \ \right]$. (See \cite{Bove1} and references
therein for a proof.)  Since (\ref{generator}) is a bounded
perturbation of $\LL^{(n)}$, it follows that
$\WW_t(\widehat{T_n})$ equals $\widehat{T_n(t)}$, where
\begin{equation}
\label{solution-n}
     T_n(t) \ = \
      \UU^{(n)}_t T_n(0) \ - \ \frac{i}{\hbar}\int_0^t \UU^{(n)}_{t-s}
                         \sum_{1 \le i< j \le n} \left[ V_{ij}, T_n(s) \right]ds
\end{equation}
when $T_n \in \TT_n$.  Taking the $m^{th}$ partial trace of both
sides of (\ref{solution-n}) yields
\begin{eqnarray*}
     T_{n:m}(t) & = &
      \UU^{(m)}_t T_{n:m}(0) \ - \ \frac{i}{\hbar}\int_0^t \UU^{(m)}_{t-s}
                       \sum_{1 \le i< j \le m} \left[ V_{ij}, T_{n:m}(s) \right] \\
        & - & (n-m)\frac{i}{\hbar}\int_0^t \UU^{(m)}_{t-s}
                      \sum_{j=1}^m \left[V_{j,m+1}, T_{n:m+1}(s) \right]_{:m}ds
\end{eqnarray*}
because of the symmetry properties of $T$ and $V$.  Multiplying
both sides of the last equation by $n!/(n-m)!$ we obtain
\begin{eqnarray}
     \frac{n!}{(n-m)!} T_{n:m}(t) & = &
      \UU^{(m)}_t \frac{n!}{(n-m)!}T_{n:m}(0) \ - \ \frac{i}{\hbar}\int_0^t \UU^{(m)}_{t-s}
              \sum_{1 \le i< j \le m}\Big[ V_{ij},\frac{n!}{(n-m)!}T_{n:m}(s) \Big] \nonumber \\
        & - & \frac{i}{\hbar}\int_0^t \UU^{(m)}_{t-s}
                 \sum_{j=1}^m \Big[V_{j,m+1}, \frac{n!}{(n-m-1)!} T_{n:m+1}(s)
                 \Big]_{:m}ds.
\label{solution-n:m}
\end{eqnarray}

Now let $D=\oplus D_n$ be a density operator on $\FFF_{\HHH}$.
Then $\WW_t D = \oplus D_n(t)$, where $D_n(t)$ is the solution of
(\ref{solution-n}) with initial condition $D_n(0)=D_n$. If $D$
satisfies moment condition (\ref{m-moment}) then
\[
     \NN_m(D(t)) \ = \  \sum_{n=m}^{\infty}\frac{n!}{(n-m)!} D(t)_{n:m}
\]
is trace class. We abbreviate $\NN_m(D(t))$ by
$\NN_m(t)$. Summing the right-hand sides of (\ref{solution-n:m})
with $D$ in place of $T$ yields
\begin{eqnarray}
     \NN_m(t) & = &
      \UU^{(m)}_t \NN_m(0) \ - \ \frac{i}{\hbar}\int_0^t \UU^{(m)}_{t-s}
              \sum_{1 \le i< j \le m}\left[ V_{ij}, \NN_m(s) \right] \nonumber \\
        & - & \frac{i}{\hbar}\int_0^t \UU^{(m)}_{t-s}
                 \sum_{j=1}^m \left[V_{j,m+1}, \NN_{m+1}(s) \right]_{:m}ds \ ,
\label{NNm(t)}
\end{eqnarray}
for (\ref{m-moment}) permits the interchange of the sum and the
integral. This is equation (\ref{vN1}) in integral form when
$m=1$. To summarize:
\begin{proposition}
\label{summarize}
 Let $\UU_t$ and $\WW_t$ be as defined in
(\ref{noninteracting}) and (\ref{dynamics}).

Suppose that $D$ is a density operator on $\FFF_{\HHH}$ of the
form $D=\oplus D_n$ such that moment condition (\ref{m-moment})
holds for some $m \in \NNN$. Let $\NN_m(t)$ denote
$\NN_m(\WW_t(D))$.  Then $\NN_m(t)$ satisfies equation
(\ref{NNm(t)}).
\end{proposition}




\section{The TDHF hierarchy}
\label{hartreehierarchy}

The existence and uniqueness of mild solutions of the TDHF
equation (\ref{TDHF}) is established in \cite{Bove1}.   There it
is shown that the integral equation
\begin{equation}
 F(t) \ = \  \UU^{(1)}_t F(0) \ - \ \frac{i}{\hbar}\int_0^t \UU^{(1)}_{t-s}
            \left[V, \ F(s)^{\otimes 2} 2A_2\right]_{:1}ds
 \label{TDHF-mild}
\end{equation}
has a unique solution $F(t)$ for any Hermitian trace class
operator $F(0)$.   Define $\FF_1(t) = F(t)$ and, for $m > 1$,
define
\begin{equation}
 \FF_m(t) \ = \  F(t)^{\otimes m}m! A_m.
\label{FFm}
\end{equation}

We proceed to derive equations for the $\FF_m(t)$ from
(\ref{TDHF-mild}).

First, set $G(t) = \UU^{(1)}_{-t} F(t)$, so that
\[
    G(t) \ = \ F(0) \ - \ \frac{i}{\hbar}\int_0^t \UU^{(1)}_{-s}
            \left[V, \ F(s)^{\otimes 2} 2A_2\right]_{:1}ds.
\]
Now apply the product rule (in integral form) to $G(t)^{\otimes
m}$:
\begin{eqnarray*}
    G(t)^{\otimes m} & = &
       F(0)^{\otimes m} \ - \ \frac{i}{\hbar}
      \sum_{j=1}^m \int_0^t  G(s)^{\otimes j-1} \otimes \UU^{(1)}_{-s}
            \left[V, \ F(s)^{\otimes 2} 2A_2\right]_{:1} \otimes  G(s)^{\otimes n-j}ds
          \\
      & = &
      F(0)^{\otimes m} \ - \ \frac{i}{\hbar}
      \int_0^t \UU^{(m)}_{-s} \sum_{j=1}^m
      \Big[V_{j,m+1}, \ F(s)^{\otimes m+1} \big(I-U_{(j,m+1)}\big)\Big]_{:m} ds .
\end{eqnarray*}
Apply $\UU^{(m)}_t$ to both sides of the preceding equation to obtain
\[
    F(t)^{\otimes m} \ = \  \UU^{(m)}_t F(0)^{\otimes m} \ - \ \frac{i}{\hbar}
    \int_0^t \UU^{(m)}_{t-s} \sum_{j=1}^m
    \Big[V_{j,m+1}, \ F(s)^{\otimes m+1} \big(I-U_{(j,m+1)}\big)\Big]_{:m} ds .
\]
Multiply both sides of the last equation by $m!A_m$ on the left,
noting that $\UU^{(m)}_s(X)A_m = \UU^{(m)}_s(XA_m)$ and that $A_m$
commutes with $\sum V_{j,m+1}$:
\[
    \FF_m(t) \ = \ \UU^{(m)}_t \FF_m(0) \ - \ \frac{i}{\hbar}
    \int_0^t \UU^{(m)}_{t-s} \sum_{j=1}^m
    \Big[V_{j,m+1}, \ F(s)^{\otimes m+1} \big(I-U_{(j,m+1)}\big)m!A_m \Big]_{:m} ds .
\]
Since $(m+1)!A_{m+1} = \big(I - U_{(1,m+1)} \  \cdots \ -
U_{(m,m+1)}\big)m!A_m$, the last equation may be rewritten
\begin{eqnarray}
    \FF_m(t) & = &
     \UU^{(m)}_t \FF_m(0) \ - \ \frac{i}{\hbar}
    \int_0^t \UU^{(m)}_{t-s} \sum_{j=1}^m
    \left[V_{j,m+1}, \ \FF_{m+1}(s)  \right]_{:m} ds  \nonumber \\
    & + &
      \sum_{1 \le j \ne k \le m} \frac{i}{\hbar}\int_0^t \UU^{(m)}_{t-s}
    \Big[V_{j,m+1}, \  F(s)^{\otimes m+1} U_{(k,m+1)}m!A_m \Big]_{:m} ds
    . \nonumber \\
\label{FFm(t)}
\end{eqnarray}
We call these equations for the $\FF_m(t)$ the {\it TDHF
hierarchy}.

 The trace norm of the
last term in (\ref{FFm(t)}) is bounded by
\[
       m(m-1)\frac{2}{\hbar} \int_0^t \Big\| \left\{ V_{m-1,m+1} U_{(m,m+1)}
       \big(\FF_m(s)\otimes F(s)\big)\right\}_{:m} \Big\|_1 ds.
\]
It can be verified that
\[
\left( V_{m-1,m+1}U_{(m,m+1)} (\FF_m(s)\otimes F(s)) \right)_{:m}
\ = \ \big(I^{\otimes m-1}\otimes F(s) \big)V_{m-1,m} \FF_m(s),
\]
whence the trace norm of the last term in (\ref{FFm(t)}) does not
exceed
\begin{equation}
\label{pre-pre-important-estimate}
       m(m-1)\frac{2}{\hbar} \int_0^t
       \|V\| \ \|F(s)\| \ \big\|\FF_m(s)\big\|_1 ds.
\end{equation}
Now $\|\FF_m(s)\|_1 \le \| F(s) \|_1^m$ by
Proposition~\ref{Prop1.5} since $\FF_m=F^{\otimes m}m!A_m$.
Furthermore, $\|F(s)\|_1 = \|F(0)\|_1$ and $\|F(s)\| = \|F(0)\|$
for all $s>0$ by Proposition~4.3 of \cite{Bove1}. Substituting
into (\ref{pre-pre-important-estimate}) produces the bound
\begin{equation}
\label{pre-important-estimate}
       m(m-1)\frac{2\|V\|}{\hbar}\|F(0)\| \ \big\|F(0)\big\|_1^m t
\end{equation}
on the trace norm of the last term in (\ref{FFm(t)}).




\section{Accuracy of the TDHF approximation}
\label{accuracy}

We have shown that if $D$ is a density operator on $\FFF_{\HHH}$ of the
form $D=\oplus D_n$ such that $\sum n^2\TR(D_n) < \infty$, then
\[
     \NN_1(t) \ = \ \UU^{(1)}_t \NN_1(0)
     \ - \  \frac{i}{\hbar}\int_0^t \UU^{(1)}_{t-s}
                  \left[V_{j,2}, \NN_{2}(s) \right]_{:1}ds \ ,
\]
where $\NN_1$ and $\NN_2$ are the one-particle and
two-particle reduced number density operators for a system which
evolved under the dynamics (\ref{Ham}) from the initial state $D$.
In this section we will compare $\NN_1(t)$ to the solution of the TDHF equation
\begin{eqnarray}
 F(t) & = &
  \UU^{(1)}_t F(0) \ - \ \frac{i}{\hbar}\int_0^t \UU^{(1)}_{t-s}
            \left[V, \ F(s)^{\otimes 2} 2A_2\right]_{:1}ds \nonumber \\
 F(0) & = & \NN_1(0)
 \label{TDHF-repeat}
\end{eqnarray}
whose initial condition is $\NN_1(0)$.
When the initial state $D$ is a Gibbs density for noninteracting fermions,
then the distance between $\NN_1(t)$ and $F(t)$ in trace norm can
be controlled at short times (Theorem~\ref{short-times}).
In the mean field limit, $F(t)$ is an asymptotically accurate
approximation to $\NN_1(t)$ at all times $t$, provided that the initial states
are Gibbs states for noninteracting fermions
(Theorem~\ref{mean-field-theorem}).

We state and discuss Theorems~\ref{short-times} and \ref{mean-field-theorem}
before going on to prove them:

\begin{theorem}
\label{short-times}
 Let $G$ be the density operator on $\FFF_{\HHH}$ of a
 Gibbs equilibrium state for noninteracting fermions, as in Proposition~\ref{Prop1}.
Let $\NN_1(t)$ and $\NN_2(t)$ denote $\NN_1(\WW_t(G))$ and $\NN_2(\WW_t(G))$,
respectively, where $\WW_t$
is the dynamics with two-particle interactions defined in (\ref{dynamics}).

Let $F(t)$ be the
solution of the TDHF equation (\ref{TDHF-repeat}).

Let $\tau$ denote $(2\|V\|\| \NN_1 \|_1)^{-1}\hbar$.
Then
\begin{equation}
\label{short-time-error}
         \big\| \NN_1(t) - F(t) \big\|_1
              \ \le \
              \frac{3}{2}
          \left( \frac{t}{\tau -  t}\right)^2
\end{equation}
for $t < \tau$.
\end{theorem}

This theorem implies, for instance, that
\[
         \big\| \NN_1(\tau/2) - F(\tau/2) \big\|_1
              \ \le \
              \frac{3}{2}.
\]
 This is remarkable because there are about $\frac{1}{2}
\|\NN_1\|_1^2$ interactions driving the dynamics (\ref{Ham}) and
the error could be much larger {\it prima
facie}: it could be proportional to $\|\NN_1\|_1$.
Unfortunately, the bound
(\ref{short-time-error}) on the error of the TDHF approximation is
valid only when $t < \tau$, which is inversely proportional to
$\| \NN_1 \|_1$,
and we have no explicit bounds for larger $t$.
If $\|\NN_1\|_1$ is too large, the bound
(\ref{short-time-error}) is useless, for then it is valid for too
short a time.

  The time-of-validity of (\ref{short-time-error})
ends up being inversely proportional to $\|\NN_1\|_1$ because the number of
two-particle interactions is proportional to $\|\NN_1\|_1^2$, and none of
these interactions is weaker than any other {\it a priori}.  In
the thermodynamic limit, where $\|\NN_1\|_1 \longrightarrow \infty$
with constant spatial density, the total interaction energy grows like $\|\NN_1\|_1$
(rather than the square of $\|\NN_1\|_1$) if the interaction potential is short-ranged.
We do not know how to derive
the TDHF equation in the thermodynamic limit, but we can derive it in the
{\it mean field} limit, where the strength of the interaction is scaled in
inverse proportion to $\|\NN_1\|_1$.

For each value of the parameter $\lambda > 0$, consider the Hamiltonian
\begin{equation}
\label{LambdaHam}
       H^{\lambda} \ = \ \sum_{i,j} \langle j| L| i \rangle a^{\dagger}_j a_i
       \ + \ \lambda\sum_{i,j,k,l} \langle k l | V | i j \rangle
       a^{\dagger}_k
       a^{\dagger}_l a_j a_i.
\end{equation}
If the initial state is given by a Gibbs equilibrium
density $D^{\lambda}$ for noninteracting fermions, then Proposition~\ref{summarize} implies
that all reduced number density operators exist and satisfy
\begin{eqnarray}
     \NN^{\lambda}_m(t) & = &
      \UU^{(m)}_t \NN^{\lambda}_m(0) \ - \
              \lambda\frac{i}{\hbar}\int_0^t \UU^{(m)}_{t-s}
              \sum_{1 \le i< j \le m}\left[ V_{ij}, \NN^{\lambda}_m(s) \right] \nonumber \\
        & - & \lambda\frac{i}{\hbar}\int_0^t \UU^{(m)}_{t-s}
                 \sum_{j=1}^m \left[V_{j,m+1}, \NN^{\lambda}_{m+1}(s) \right]_{:m}ds \nonumber \\
      \NN^{\lambda}_m(0) & = & \NN_m(D^{\lambda}(0)) .
\label{LambdaNNm}
\end{eqnarray}
We will show that if $\lambda$ is inversely
proportional to $\|\NN^{\lambda}_1\|_1$, then $\NN^{\lambda}_m(t)$
is close to $F^{\lambda}(t)^{\otimes m}m! A_m$ in trace norm, where
\begin{eqnarray}
 F^{\lambda}(t) & = &  \UU^{(1)}_t F^{\lambda}(0)
            \ - \ \frac{i}{\hbar}\int_0^t \UU^{(1)}_{t-s}
            \left[V, \ F^{\lambda}(s)^{\otimes 2} 2A_2\right]_{:1}ds
        \nonumber \\
 F^{\lambda}(0) & = & \NN^{\lambda}_1(0).
 \label{TDHF-lambda}
\end{eqnarray}

\begin{theorem}
\label{mean-field-theorem}
Let $\left\{D^{\lambda}\right\}_{\lambda >0}$ be a
family of Gibbs equilibrium densities for
noninteracting fermions, as in Proposition~\ref{Prop1}, with
\[
\limsup\limits_{\lambda \rightarrow 0} \lambda \|\NN_1(D^{\lambda})\|_1
     \ < \  \infty.
\]
Let $\NN^{\lambda}_m(t)$ be the solution
of (\ref{LambdaNNm}), let $F^{\lambda}(t)$ be the
solution of the TDHF equation
(\ref{TDHF-lambda}) and let $\FF^{\lambda}_m(t) =
F^{\lambda}(t)^{\otimes m}m! A_m$.
 Then
\[
      \lim_{\lambda \rightarrow 0}
      \big\| \NN^{\lambda}_m(t) - \FF^{\lambda}_m(t)
      \big\|_1 \big/ \|\NN^{\lambda}_1\|_1^{m}
      \ = \ 0
\]
for all $t>0$ and all $m \in \NNN$.
\end{theorem}

We will derive
Theorems~\ref{short-times} and \ref{mean-field-theorem}
from Lemma~\ref{full-short-time} below.

Let $\NN_m(t)$ be as in Proposition~\ref{summarize}, and let
$\FF_m(t)$ satisfy the TDHF hierarchy.  In the hypotheses
Theorems~\ref{short-times} and \ref{mean-field-theorem} we
suppose that $F(0)=\NN_1(D(0))$, but for now let us only assume
that
\begin{equation}
\label{boundsOnF}
     \|F(0)\|\le 1 \qquad \hbox{and} \qquad \|F(0)\|_1 =
     \|\NN_1(D(0))\|_1.
\end{equation}
The trace norm of $\NN_1(t)$ is independent of $t$, and we shall
denote it simply by $\|\NN_1 \|_1$.
 Assuming (\ref{boundsOnF}), the
bound (\ref{pre-important-estimate}) is itself bounded by
\begin{equation}
\label{important-estimate}
       m(m-1)\frac{2\|V\|}{\hbar} \ \big\|\NN_1 \big\|_1^m t.
\end{equation}
 Subtracting equations (\ref{FFm(t)}) from
equations (\ref{NNm(t)}) and using (\ref{important-estimate})
leads to the estimates
\begin{eqnarray*}
             \big\| \NN_m(t) - \FF_m(t) \big\|_1
         & \le &
         \big\| \NN_m(0) - \FF_m(0) \big\|_1  \\
         & + &
          \frac{m(m-1)}{\hbar}\|V\|
                 \Big( 2 \| \NN_1 \|_1^m +  \| \NN_m \|_1 \Big)t  \\
         & + &
         m\frac{2 \|V\|}{\hbar}
                \int_0^t \big\| \NN_{m+1}(s) - \FF_{m+1}(s) \big\|_1 ds \ .
\end{eqnarray*}
Iterating this estimate $n$ times, one obtains
\begin{eqnarray}
             \big\| \NN_m(t) - \FF_m(t) \big\|_1
         & \le &
         \sum_{j=0}^{n} a_{m+j} \binom{m+j-1}{j}   C^j t^j
         \ + \
         \sum_{j=0}^{n} \frac{b_{m+j}}{j+1}
        \binom{m+j-1}{j}   C^j t^{j+1}
         \nonumber \\
         & + &
         C^n \frac{(m+n)!}{(m-1)!}
          \int_0^t \int_0^{t_1} \cdots \int_0^{t_{n}}
            \big\| \NN_{m+n+1}(s) - \FF_{m+n+1}(s) \big\|_1
        ds dt_n \cdots dt_1 \nonumber \\
\qquad \label{key-estimate}
\end{eqnarray}
with $ C = 2\|V\|/\hbar$ and
\begin{eqnarray}
        a_m & = & \big\| \NN_m(0) - \FF_m(0) \big\|_1 \nonumber \\
    b_m & = & \frac{m(m-1)}{\hbar}\|V\|
                 \Big( 2 \| \NN_1 \|_1^m  +  \| \NN_m \|_1 \Big).
\label{key-notation}
\end{eqnarray}
To make use of these estimates we need some control over the size of the integrand in
(\ref{key-estimate}).
We will assume that there exists a constant $B$ such that
\begin{equation}
       \| \NN_m \|_1 \ \le \  B\| \NN_1 \|_1^m
\label{moment-hypothesis}
\end{equation}
for all $m$.  Then the last term on the right hand side of
(\ref{key-estimate}) is bounded by
\[
    \binom{m+n-1}{m-1}(B+1)\big\| \NN_1 \big\|_1^{m+n}C^n
    t^n ,
\]
which tends to $0$ as $n$ tends to infinity if $m$ fixed and
$C\|\NN_1\|_1t < 1$. Furthermore, assuming
(\ref{moment-hypothesis}), we can bound $b_m$ of
(\ref{key-notation}) by $ C(B/2+1)\| \NN_1 \|_1^m m(m-1)$ and we
find the following:
\begin{lemma}
\label{full-short-time}
 Suppose that $D$ is a density operator on
$\FFF_{\HHH}$ of the form $D=\oplus D_n$, such that
(\ref{moment-hypothesis}) holds for all $m \in \NNN$.  Let $\WW_t$
be as defined in (\ref{dynamics}) and let $\NN_m(t)$ denote
$\NN_m(\WW_t(D))$. Let $F(t)$ be the solution of a TDHF equation
(\ref{TDHF-mild}) whose initial condition $F(0)$ satisfies
(\ref{boundsOnF}), and let $\FF_m(t)$ be as in (\ref{FFm}). Then,
with $C=2\|V\|/\hbar$,
\begin{eqnarray*}
            \frac{ \big\| \NN_m(t) - \FF_m(t) \big\|_1 }{\big\| \NN_1 \big\|_1^m}
               & \le &
         \sum_{j=0}^\infty
         \frac{\big\| \NN_{m+j}(0) - \FF_{m+j}(0) \big\|_1 } {\big\| \NN_1 \big\|_1^{m+j} }
          \binom{m+j-1}{m-1}
           \big( C \| \NN_1 \|_1  t\big)^j \\
           & + &
          \big\| \NN_1 \big\|_1^{-1} \frac{B+2}{2}
              \sum_{j=0}^\infty (m+j-1) \binom{m+j}{m-1}
         \big( C \| \NN_1 \|_1  t\big)^{j+1}
\end{eqnarray*}
when $ C \| \NN_1 \|_1  t <1 $.
\end{lemma}

\noindent {\bf Proof of Theorem~\ref{short-times}}
When $G$ is the
density operator of a Gibbs equilibrium state and
$F(0)=\NN_1(G)$, then
  $\|F(0)\| = \|\NN_1(G)\| \le 1$ by Proposition~\ref{N1},
$\big\| \NN_m(0) - \FF_m(0) \big\|_1 = 0$ for all $m$ by
Proposition~\ref{Prop1}, and $B=1$
in (\ref{moment-hypothesis}) by Proposition~\ref{Prop2}.
Upon simplifying the inequality in Lemma~\ref{full-short-time}, one
obtains Theorem~\ref{short-times}.
\hfill $\square$

\noindent {\bf Proof of Theorem~\ref{mean-field-theorem}}
For $\NN^{\lambda}_m(t)$ and $\FF^{\lambda}_m(t)$ as in the hypothesis of
Theorem~\ref{mean-field-theorem}, observe that
\[
     \|F(t)\|\le 1 \qquad \hbox{and} \qquad \|F(t)\|_1 =
     \|\NN_1(D(t))\|_1
\]
at all times $t$, and $\NN^{\lambda}_m$ satisfies
(\ref{moment-hypothesis}) with  $B=1$.  Thus, we may apply
Lemma~\ref{full-short-time}, with a few changes:  $t$ and
$t+\Delta t$ may be substituted for $0$ and $t$, and $C$ should be
replaced by $C\lambda$.  These substitutions yield
\begin{eqnarray}
            \frac{ \big\| \NN^{\lambda}_m(t+\Delta t) -
        \FF^{\lambda}_m(t+\Delta t) \big\|_1 }{\|\NN^{\lambda}_1\|_1^m}
               & \le &
         \sum_{j=0}^\infty
         \frac{\big\| \NN^{\lambda}_{m+j}(t) -
     \FF^{\lambda}_{m+j}(t) \big\|_1 } {\|\NN^{\lambda}_1\|_1^{m+j} }
          \binom{m+j-1}{m-1}
           \big( C \lambda \|\NN^{\lambda}_1\|_1 \Delta t \big)^j \nonumber \\
           & + &
          \frac{1}{\|\NN^{\lambda}_1\|_1} \frac{3}{2}
              \sum_{j=0}^\infty (m+j-1) \binom{m+j}{m-1}
         \big( C \lambda \|\NN^{\lambda}_1\|_1 \Delta t \big)^{j+1} \nonumber \\
\label{take-limit}
\end{eqnarray}
for $\Delta t < (C\lambda \|\NN^{\lambda}_1\|_1)^{-1}$.
Since $
      u =  \limsup\limits_{\lambda\rightarrow 0}
      C \lambda \|\NN^{\lambda}_1\|_1
$
is finite by hypothesis, taking the $\limsup$ of both sides of (\ref{take-limit})
implies that
\begin{equation}
\label{Slater-closure}
      \lim_{\lambda\rightarrow 0}  \big\| \NN^{\lambda}_m(s)
      - \FF^{\lambda}_m(s)  \big\|_1
      \big/
       \|\NN^{\lambda}_1\|_1^{m}
      \ = \ 0 \qquad \forall \ m \in \NNN
\end{equation}
holds at time $s=t+\Delta t$ if it holds at time $s=t$ and $\Delta
t < 1/u$. Since (\ref{Slater-closure}) holds at $s=0$, an
inductive argument proves that it holds at all times $s>0$.
\hfill $\square$




\section{Appendix: the proofs of Propositions~\ref{N1},
\ref{Prop1}, and \ref{Prop1.5}}

\subsection{Proof of Proposition~\ref{N1}}
 We begin by proving
\begin{proposition}
\label{ColemanProp} If $D_n$ is an n-particle fermionic density
operator then
\begin{equation}
\label{Coleman}
                \big\| D_{n:1} \big\| \ \le \ 1/n.
\end{equation}
\end{proposition}
\noindent {\bf Proof}: \qquad Thanks to the convexity of the norm
and the linearity of the partial trace, it suffices to prove
(\ref{Coleman}) for fermionic pure states.  Let $\HHH$ denote the
single-particle Hilbert space, and let $\Psi$ be a unit vector in
$\HHH^{(n)}$. Since $P_{\Psi:1}$ is a compact Hermitian operator,
there exists a unit vector $u \in \HHH$ such that
\begin{equation}
\label{PfEqA}
             \big\|P_{\Psi:1} \big\| \ = \
             \langle u,P_{\Psi:1}(u)\rangle.
\end{equation}
Let $\{\phi_j\}_{j \in J}$ be a basis of $\HHH$ containing $u$.
For each subset ${\bf s}=\{ j_1,\ldots, j_n\}$ of $J$, let
$\Psi({\bf s})$ denote one of the two Slater determinants that may
be formed from the vectors $\phi_{j_1},\ldots,\phi_{j_n}$ (the two
choices differ only in sign). The set of vectors $\Psi({\bf s})$
is an orthonormal basis of $\HHH^{(n)}$ and so $ \Psi = \sum_{{\bf
s}} \langle \Psi({\bf s}),\Psi
        \rangle \Psi({\bf s})$.
By definition of the partial trace, $\langle
u,P_{\Psi:1}(u)\rangle$ equals
\begin{equation}
\label{PfEqB}
          \sum_{j_1,\ldots,j_{n-1} \in J} \Big\langle u \otimes \phi_{j_1} \otimes \cdots \otimes \phi_{j_{n-1}},
          \ P_{\Psi}(u \otimes \phi_{j_1}  \otimes \cdots \otimes \phi_{j_{n-1}}) \Big\rangle.
\end{equation}
From (\ref{PfEqA}) and (\ref{PfEqB})
\begin{eqnarray}
\label{PfEqC}
       \left\|P_{\Psi:1} \right\|
          & = &
          \sum_{j_1,\ldots,j_{n-1} \in J} \left\| P_{\Psi}(u \otimes \phi_{j_1}
           \otimes \cdots \otimes \phi_{j_{n-1}}) \right\|^2
          \nonumber \\
          & = &
          \sum_{j_1,\ldots,j_{n-1} \in J} \sum_{\bf s} \left|\langle \Psi({\bf s}),\Psi
        \rangle\right|^2
        \left|
          \big\langle \Psi({\bf s}), u \otimes \phi_{j_1} \otimes \cdots \otimes \phi_{j_{n-1}}
          \big\rangle\right|^2. \nonumber \\
\end{eqnarray}
But
\[
     \left|\big\langle \Psi({\bf s}), u \otimes \phi_{j_1} \otimes
          \cdots \otimes \phi_{j_{n-1}} \big\rangle\right|^2
\ = \ \bigg\{
      \begin{array}{cl}
           1/n! \  & \hbox{if} \quad {\bf s} = \{u,\phi_{j_1},\ldots,\phi_{j_{n-1}}\} \\
           0 \  & \hbox{otherwise} \\
      \end{array}
\]
whence
\[
          \left\|P_{\Psi:1} \right\|
          \ \le \
            \sum_{\bf s} \left|\langle \Psi({\bf s}),\Psi \rangle\right|^2
            \frac{(n-1)!}{n!}
          \ = \ \frac{1}{n}\sum_{\bf s} \left|\langle \Psi({\bf s}),\Psi \rangle\right|^2 \ = \ \frac{1}{n}
\]
by (\ref{PfEqC}). \hfill $\square$

Now we return to the proof of Proposition~\ref{N1}.

 Let $D$ be a density operator on
$\FFF_{\HHH}$ that commutes with the number operator $N$ and
satisfies $\TR(ND) < \infty$.  Then $D$ has the form
(\ref{directsum}) and $\sum n \TR(D_n) < \infty$, so that
$\NN_1(D)= \sum nD_{n:1}$ is defined and
\[
        \big\| \NN_1(D) \big\| \ \le \ \sum_{n=1}^{\infty} n \big\| D_{n:1}
        \big\|.
\]
By Proposition~\ref{ColemanProp} the operator norm of $D_n$ is
less than or equal to $\TR(D_n)/n$, whence
\[
        \big\| \NN_1(D) \big\| \ \le \ \sum_{n=1}^{\infty} n \TR(D_n)/n \
        = \ \TR(D) \ = \ 1.
\]

\subsection{Proof of Proposition~\ref{Prop1}}

  From (\ref{G}), with $ \PPP ({\bf s})$ as in
  (\ref{Prob}),
\[
    \NN_n(G)
      \ = \
     \sum_{{\bf s}: N({\bf s}) \ge n} \frac{N({\bf s})!}{(N({\bf s})-n)!}
     \PPP ({\bf s})\left( P_{\Psi({\bf s})}\right)_{:n}.
\]
 Substituting the
expressions (\ref{SlaterFormula}) and (\ref{SlaterOne}) for
$\left( P_{\Psi({\bf s})}\right)_{:n}$ and collecting terms, we
find that
\begin{eqnarray}
     \NN_n(G)
      & = &
     \sum_{{\bf s}: N({\bf s}) \ge n}
     \PPP ({\bf s})\left( N({\bf s}) P_{\Psi(\bf{s})}\right)_{:1}^{\ \otimes n}
      n!A_n \nonumber \\
     & = &
     \sum_{\stackrel{distinct}{j_1,\ldots, j_n \in J}}
     \Big[
     \sum_{{\bf s} \ni j_1,\ldots, j_n }
             \PPP ({\bf s})
     \Big]
     \big(
     P_{\phi_{j_1}}\otimes \cdots \otimes P_{\phi_{j_n}}
     \big)
    n!A_n \ , \label{Gn}
\end{eqnarray}
where $N({\bf s})$ is the size of ${\bf s}$. The sum in (\ref{Gn})
is made over {\it distinct} $j_1,\ldots,j_n$ since
$(P_{\phi_{j_1}}\otimes \cdots \otimes P_{\phi_{j_n}} ) A_n$
equals the zero operator if $j_r = j_s$ for any $r \ne s$. In
particular,
\begin{equation}
\label{a}
     \NN_1(G)
     \ = \
     \sum_{j \in J}
     \Big[ \sum_{{\bf s} \ni j}
             \PPP ({\bf s})\Big]
     P_{\phi_j}
    \ = \
         \sum_{j \in J}
     p(j)
     P_{\phi_j}.
\end{equation}
Thus,
\begin{equation}
\label{b}
 \NN_1(G)^{\otimes n}n!A_n
 \ = \
  \sum_{\stackrel{distinct}{j_1,\ldots, j_n \in J}}
    \prod_{i=1}^n p (j_i)
       \big(  P_{\phi_{j_1}}\otimes \cdots \otimes P_{\phi_{j_n}}\big) n! A_n .
\end{equation}
The sum in (\ref{b}) is again restricted to distinct
$j_1,\ldots,j_n$ because of the presence of the antisymmetrizer
 $A_n $.  Substituting
(\ref{useMe}) into (\ref{b}) yields (\ref{Gn}), proving the
proposition.

\subsection{Proof of Proposition~\ref{Prop1.5}}

 Suppose $T$ is a Hermitian trace
class operator.  There exists an orthonormal basis $\{ e_j
\}_{j\in J}$ of $\HHH$ such that
\[
     T \ = \ \sum_j \lambda_j P_{e_j} \qquad \hbox{with} \qquad
     \sum_j |\lambda_j| \ = \ \big\| T \big\|_1.
\]
The operator $T^{\otimes n}n!A_n$ is diagonalizable with respect
to the basis of Slater determinants formed from $n$ distinct
members of $\{ e_j \}$.  Indeed,
\begin{eqnarray*}
     T^{\otimes n}n!A_n \big( \sqrt{n!} \ A_n(e_{j_1} \otimes \cdots \otimes
     e_{j_n})\big) & = & \sqrt{n!} A_n n! T^{\otimes n}(e_{j_1} \otimes  \cdots \otimes
     e_{j_n})  \\
     & = &
     \Big( n! \prod_{s=1}^n \lambda_{j_s}
     \Big) \sqrt{n!} \ A_n(e_{j_1} \otimes  \cdots \otimes
     e_{j_n}).
\end{eqnarray*}
The trace norm of $T^{\otimes n}n!A_n$ is the sum of the absolute
values of its eigenvalues, whence
\begin{equation}
\label{traceNormOfAntisymmetrizedPower}
    \big\| T^{\otimes n}n!A_n  \big\|_1 \ = \  \sum_{\stackrel{distinct}{j_1,\ldots, j_n \in J}}  \prod_{s=1}^n \left|
     \lambda_{j_s}\right|.
\end{equation}
Note that the sum in (\ref{traceNormOfAntisymmetrizedPower}) is
over ordered sequences $j_1,j_2,\ldots, j_n$ rather than subsets
$\{j_1,\ldots, j_n\}$.  But
\[
\sum_{\stackrel{distinct}{j_1,\ldots, j_n \in J}}  \prod_{s=1}^n
\left|
     \lambda_{j_s}\right| \ \le \ \sum_{j_1,\ldots, j_n \in J}  \prod_{s=1}^n \left|
     \lambda_{j_s}\right| \ = \ \prod_{s=1}^n \sum_{j \in J}
     |\lambda_j| \ = \  \| T \|_1^n
\]
proving (\ref{proveMe}).




\end{document}